\documentclass[conference]{IEEEtran}
\IEEEoverridecommandlockouts

\usepackage{cite}
\usepackage{amsmath,amssymb,amsfonts}
\usepackage{algorithmic}
\usepackage{graphicx}
\usepackage{textcomp}
\usepackage{xcolor}

\def\BibTeX{{\rm B\kern-.05em{\sc i\kern-.025em b}\kern-.08em
    T\kern-.1667em\lower.7ex\hbox{E}\kern-.125emX}}
\begin{document}

\title{From DeepSense to Open RAN: AI/ML Advancements in Dynamic Spectrum Sensing and Their Applications\\
}

\author{\IEEEauthorblockN{Ryan Barker, Tolunay Seyfi, Fatemeh Afghah} \\
    \IEEEauthorblockA{Holcombe Department of Electrical and Computer Engineering, \\
    Clemson University, Clemson, SC, USA \\
    Emails: \{rcbarke, tseyfi, fafghah\}@clemson.edu}
}

\maketitle

\begin{abstract}
The integration of Artificial Intelligence (AI) and Machine Learning (ML) in next-generation wireless communication systems has become a cornerstone for advancing intelligent, adaptive, and scalable networks. This reading report examines key innovations in dynamic spectrum sensing (DSS), beginning with the foundational DeepSense framework, which uses convolutional neural networks (CNNs) and spectrogram-based analysis for real-time wideband spectrum monitoring. Building on this groundwork, it highlights advancements such as DeepSweep and Wideband Signal Stitching, which address the challenges of scalability, latency, and dataset diversity through parallel processing, semantic segmentation, and robust data augmentation strategies. The report then explores Open Radio Access Networks (ORAN), focusing on AI/ML-driven enhancements for UAV experimentation, digital twin-based optimization, network slicing, and self-healing xApp development. By bridging AI-based DSS methodologies with ORAN’s open, vendor-neutral architecture, these studies underscore the potential of software-defined, intelligent infrastructures in enabling efficient, resilient, and self-optimizing networks for 5G/6G ecosystems. Through this synthesis, the report highlights AI’s transformative role in shaping the future of wireless communication and autonomous systems.
\end{abstract}

\begin{IEEEkeywords}
Artificial Intelligence (AI), Machine Learning (ML), Dynamic Spectrum Sensing (DSS), Open Radio Access Networks (ORAN), DeepSense, DeepSweep, Wideband Signal Stitching, Convolutional Neural Networks (CNNs), Semantic Spectrum Segmentation, Digital Twins, xApps, UAV Experimentation, Network Slicing, 5G, 6G, Edge Computing, Real-time Spectrum Analysis.
\end{IEEEkeywords}

\section{Introduction}
The landscape of wireless communication is undergoing a transformative evolution, driven by the convergence of Artificial Intelligence (AI), Machine Learning (ML), and open architectural paradigms. As the demand for scalability, efficiency, and flexibility intensifies in the era of 5G and beyond, technologies like DeepSense and Open Radio Access Networks (ORAN) are leading this revolution. These innovations have emerged as pivotal enablers of real-time, adaptive, and intelligent networking solutions, addressing the growing complexities of next-generation systems.

\begin{figure}[htbp] \centering \includegraphics[width=\columnwidth]{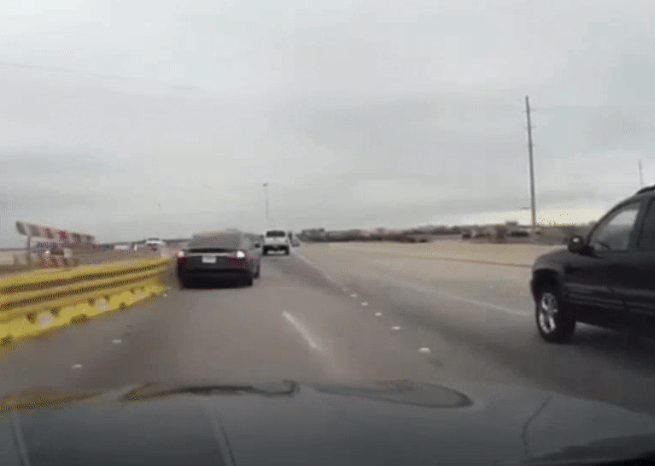} \caption{Example of latency-sensitive applications in autonomous vehicles. Delays in spectrum sensing can lead to catastrophic outcomes, such as missed collision alerts or disrupted V2V communication.} \label{fig:autonomous_crash} \end{figure}

Dynamic Spectrum Sensing (DSS) is particularly critical for latency-sensitive applications such as autonomous vehicles and ultra-reliable low-latency communication (URLLC) devices. These systems rely on uninterrupted, real-time connectivity to ensure safe and efficient operation. For instance, an autonomous vehicle navigating a busy intersection depends on instantaneous spectrum sensing to maintain reliable communication with nearby vehicles, infrastructure, and edge computing nodes. High latency in spectrum sensing could lead to delays in receiving critical data, such as collision alerts or dynamic route adjustments, resulting in catastrophic failures. A striking example of this is illustrated in Fig. \ref{fig:autonomous_crash}, where delays in communication and processing can lead to accidents on congested roads. Similarly, in industrial automation, spectrum sensing delays could disrupt machine coordination, leading to operational inefficiencies or even accidents.

The DeepSense framework represents a novel application of AI to spectrum sensing, leveraging deep learning algorithms to achieve real-time, wideband spectrum analysis. By integrating AI directly into radio frequency (RF) applications, DeepSense has addressed longstanding challenges in computational complexity and scalability, setting a foundational standard for subsequent advancements [1]. Building on this groundwork, recent studies have expanded the framework’s capabilities through methods such as semantic spectrum segmentation and scalable spectrum sensing using convolutional neural networks (CNNs). These enhancements have demonstrated significant improvements in spectrum utilization and performance across diverse wireless environments [2][3].

Simultaneously, ORAN is transforming traditional radio access networks (RANs) by introducing modular, open, and vendor-neutral design principles. This architecture fosters interoperability and innovation, enabling the deployment of intelligent applications like xApps and rApps that optimize network adaptability and resource management [6][7]. Through integration with AI/ML technologies, ORAN supports capabilities such as real-time decision-making, predictive analytics, and proactive optimization of network operations. Experimental platforms like AERPAW validate ORAN’s potential, demonstrating its adaptability in dynamic use cases such as UAV communication and digital twin simulations [6][8][10].

This report provides a comprehensive analysis of these advancements, examining foundational studies like DeepSense, iterative improvements such as DeepSweep and Wideband Signal Stitching, and ORAN’s AI-driven applications. By synthesizing insights from these innovations, the report underscores the transformative role of AI/ML and open architectures in addressing the challenges of next-generation communication systems.

\section{RELATED WORK}
The integration of Artificial Intelligence (AI) into spectrum sensing and wireless communication has catalyzed significant advancements, transforming the way networks operate and adapt. Among these contributions, the DeepSense framework [1] stands out as a pioneering solution, leveraging deep learning for real-time, wideband spectrum sensing. By employing in-the-loop deep learning, DeepSense enhances spectrum utilization by dynamically adapting to changing RF conditions, providing more accurate RF analysis and establishing a strong foundation for AI-augmented systems in wireless communication.

Building on this groundwork, subsequent studies have addressed the limitations of DeepSense while expanding its capabilities. For instance, "Stitching the Spectrum" [2] introduces semantic segmentation to create continuous representations of fragmented RF signals, improving interference identification and signal overlap resolution. Meanwhile, "DeepSweep" [3] employs a parallelized architecture using convolutional neural networks (CNNs) to enhance throughput in dynamic environments. These advancements collectively demonstrate the potential of AI to address the scalability and latency challenges of modern spectrum sensing.

Foundational studies such as "Big Data Goes Small" [4] have further highlighted the utility of deploying deep learning at the embedded systems level. By incorporating real-time feedback loops, this work showcases how ML-powered systems can optimize RF sensing in resource-constrained environments. Similarly, "DeepLab" [5], originally designed for semantic image segmentation, has been creatively adapted to spectrum analysis, demonstrating how advanced computer vision techniques can be applied to understand complex RF environments and improve spectrum management.

Parallel to these advancements, Open Radio Access Networks (ORAN) have emerged as a transformative architecture in wireless communication. ORAN’s modular and vendor-neutral design fosters interoperability and innovation, enabling rapid deployment of intelligent control applications like xApps and rApps. For example, the AERPAW testbed [6] explores how ORAN-based architectures can be adapted for real-world dynamic environments such as UAV communication. Furthermore, frameworks like xDevSM [7] streamline xApp development, showcasing their potential to enhance RAN adaptability and resource management.

Recent research has also focused on operational challenges within ORAN, such as optimizing virtual network functions (VNFs) and addressing vulnerabilities associated with AI/ML integration. "Optimizing Virtual Network Function Splitting in Open-RAN Environments" [9] presents strategies for improving resource efficiency, while "Misconfiguration in O-RAN: Analysis of the Impact of AI/ML" [10] emphasizes the need for robust testing frameworks to ensure network reliability. Together, these works highlight the importance of combining AI/ML-driven solutions with ORAN’s flexible architecture to create resilient, scalable, and efficient wireless networks.

\section{Dynamic Spectrum Sensing and AI Innovations}

\subsection{Foundational Advances in Spectrum Sensing}
Dynamic spectrum sensing (DSS) has emerged as a critical enabler for efficient spectrum utilization, especially in congested and dynamic wireless environments. The foundational DeepSense framework \cite{Khalid2020} represents a transformative approach, bridging breakthroughs in computer vision with wireless networking by leveraging convolutional neural networks (CNNs) for real-time, wideband spectrum analysis. Unlike traditional spectrum sensing methods, which rely on protocol-specific algorithms, DeepSense processes raw in-phase and quadrature (I/Q) signal data directly, bypassing the need for extensive preprocessing or reliance on feature extraction pipelines at the base station.

\begin{figure}[htbp]
    \centering
    \includegraphics[width=\columnwidth]{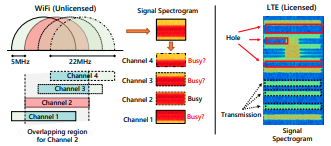}
    \caption{Spectrograms: A visual representation of signal frequency over time, illustrating their role in transforming wireless spectrum data into formats CNNs can analyze.}
    \label{fig:spectrogram_example}
\end{figure}

This ability to work directly with raw I/Q samples is a significant advancement. I/Q data captures the full amplitude and phase information of RF signals, providing a high-fidelity representation of the wireless environment. By processing this data in its raw form, DeepSense eliminates the latency and computational overhead associated with converting I/Q samples into spectrograms or other intermediate representations at the base station. Spectrograms—visual representations of signal frequency over time—are critical for enabling CNNs to analyze spectral patterns and detect interference or underutilized frequencies with high precision, as shown in Fig. \ref{fig:spectrogram_example}.

\begin{figure}[htbp]
    \centering
    \includegraphics[width=\columnwidth]{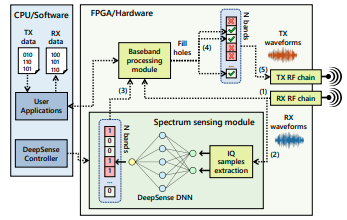}
    \caption{DeepSense System Model: Overview of the DeepSense architecture, which processes raw I/Q samples to perform real-time spectrum sensing.}
    \label{fig:deepsense_model}
\end{figure}

The DeepSense System Model, shown in Fig. \ref{fig:deepsense_model}, illustrates the framework’s architecture. It includes an integrated deep neural network (DNN) for spectrum sensing that processes raw I/Q samples, enabling real-time spectrum analysis without the need for extensive preprocessing. This streamlined design ensures that DeepSense can operate efficiently in dynamic environments.

DeepSense addresses key challenges, including the computational complexity of processing entire spectrum bands, granularity limitations in traditional approaches, and the inability of static methods to adapt to dynamic RF conditions. By leveraging CNNs originally designed for computer vision, DeepSense adapts these strengths to the wireless domain, laying the groundwork for future advancements in AI-driven spectrum sensing.

\subsection{Advanced Techniques and Iterative Improvements}
Building on DeepSense, innovations such as DeepSweep \cite{Khalid2020c} and Wideband Signal Stitching \cite{Khalid2020b} address limitations in scalability, latency, and dataset diversity.

\begin{figure}[htbp]
    \centering
    \includegraphics[width=\columnwidth]{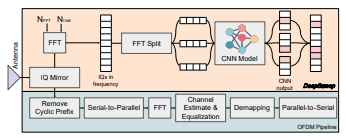}
    \caption{DeepSweep System Model: Parallelized architecture for scalable and low-latency spectrum sensing, dividing the spectrum into smaller chunks for simultaneous processing.}
    \label{fig:deepsweep_model}
\end{figure}

\begin{figure}[htbp]
    \centering
    \includegraphics[width=\columnwidth]{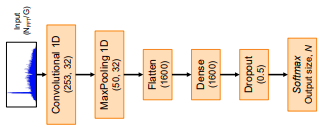}
    \caption{DeepSweep Spectrum Sensing CNN: The architecture utilizes lightweight convolutional layers, reducing complexity while enabling parallel spectrum chunk processing.}
    \label{fig:deepsweep_cnn}
\end{figure}

\textbf{DeepSweep} introduces a parallelized architecture that divides the spectrum into smaller, manageable chunks, which are processed simultaneously using lightweight CNNs. This approach achieves real-time inference with a latency of less than 1 millisecond, making it ideal for applications like autonomous vehicle networks that require rapid spectrum analysis to maintain reliable connectivity. DeepSweep’s streamlined architecture reduces computational demands by 10$\times$ while maintaining a detection accuracy of 98\% for narrowband interference, as shown in Fig. \ref{fig:deepsweep_model} and Fig. \ref{fig:deepsweep_cnn}.

\begin{figure}[htbp]
    \centering
    \includegraphics[width=\columnwidth]{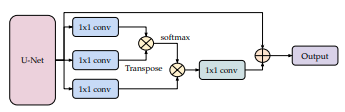}
    \caption{Semantic Spectrum Segmentation Model: Wideband Signal Stitching adapts U-Net architecture for precise signal classification and segmentation.}
    \label{fig:semantic_segmentation}
\end{figure}

\begin{figure}[htbp]
    \centering
    \includegraphics[width=\columnwidth]{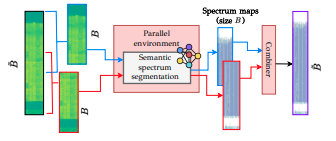}
    \caption{Scalable and Portable Pipeline: Wideband Signal Stitching’s architecture processes large bandwidths using parallel semantic segmentation environments.}
    \label{fig:scalable_pipeline}
\end{figure}

\textbf{Wideband Signal Stitching} complements DeepSweep by addressing granularity limitations and enhancing dataset diversity. Using semantic spectrum segmentation at the I/Q sample level, it bypasses pixel-based bounding boxes, which are prone to misclassifications in overlapping signal conditions. Wideband Signal Stitching introduces non-local blocks into CNNs, improving spatial feature integration for complex signal classification. Its innovative dataset generation pipeline combines over-the-air (OTA) signals with synthetic interference and noise, ensuring effective generalization across diverse RF environments \cite{Chen2018}. Figures \ref{fig:semantic_segmentation} and \ref{fig:scalable_pipeline} highlight the segmentation model and scalable pipeline used in this framework.

\subsection{Significance of DeepSense in Real-World Scenarios}
DeepSense and its iterative advancements address critical gaps in traditional spectrum sensing methodologies. By processing raw I/Q samples and adapting CNNs for spectrogram analysis, these frameworks provide unparalleled precision and adaptability.

\begin{figure}[htbp]
    \centering
    \includegraphics[width=\columnwidth]{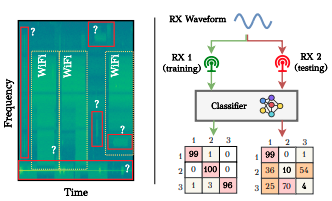}
    \caption{Challenges in Labeling Spectrum: Demonstrating the difficulty of annotating spectrum data due to overlapping and diverse wireless signals.}
    \label{fig:labeling_spectrum}
\end{figure}

\begin{enumerate}
    \item \textbf{Latency-Sensitive Applications:} For autonomous vehicles, delays in spectrum sensing can lead to catastrophic outcomes such as missed collision alerts or disruptions in vehicle-to-vehicle (V2V) communication. Similarly, in industrial automation, where machine coordination depends on seamless communication, real-time interference mitigation prevents costly downtime or accidents.
    
    \item \textbf{Scalability in Dynamic Environments:} Traditional methods often fail to scale in environments with fluctuating interference patterns or heterogeneous signal types. DeepSense, coupled with advancements like DeepSweep and Wideband Signal Stitching, adapts dynamically to changing conditions, ensuring reliable operation across scenarios such as urban IoT networks or UAV communication.
    
    \item \textbf{Improved Spectrum Efficiency:} The ability to identify spectral holes with high precision optimizes resource allocation, maximizing spectrum usage and minimizing interference.
\end{enumerate}

\subsection{Path Forward for AI-Driven Spectrum Sensing}
The evolution from DeepSense to DeepSweep and Wideband Signal Stitching exemplifies the potential of AI to revolutionize spectrum sensing. Future research should explore deeper integration of reinforcement learning for autonomous optimization of spectrum utilization. Additionally, enhancing the scalability of CNN architectures to handle terahertz frequencies and incorporating edge computing capabilities for distributed spectrum analysis could further expand the applicability of these frameworks. These advancements will be instrumental in meeting the demands of increasingly heterogeneous and latency-sensitive wireless ecosystems, paving the way for intelligent, self-optimizing communication networks.

\section{\textbf{ORAN: Digital Twins, Experimentation, and Intelligent Applications}} \label{sec:4}

\subsection{ORAN and Digital Twin Experimentation}
The Open Radio Access Network (ORAN) architecture is redefining wireless communication by enabling greater adaptability and reducing vendor dependency. A key innovation within ORAN is the concept of digital twins—virtual replicas of physical networks that allow comprehensive experimentation and optimization. Wang et al. \cite{Wang2021} highlight how digital twins facilitate the optimization of Virtual Network Function (VNF) splitting in Open-RAN environments. By simulating complex network topologies and processes, digital twins provide network operators with a controlled environment to make informed resource allocation decisions based on realistic, dynamically generated scenarios.

Digital twins also play a pivotal role in evaluating and enhancing the resilience of ORAN systems to potential misconfigurations. Allen et al. \cite{Allen2021} emphasize the use of digital twins to detect, analyze, and mitigate risks without disrupting live network operations. This pre-deployment testing framework ensures that AI/ML integration does not inadvertently compromise network reliability. By identifying vulnerabilities and offering mitigation strategies, digital twins significantly enhance the stability and security of ORAN environments.

Advancements in the physical layer of ORAN are further supported by tools like Sionna, an open-source library designed for next-generation physical layer research. Hyodis et al. \cite{Perovic2022} detail how Sionna enables researchers to model physical processes such as beamforming and signal propagation, ensuring that ORAN’s logical control mechanisms align seamlessly with physical network conditions. This integration contributes to more robust and efficient ORAN deployments, paving the way for future innovations in network architecture.

\begin{figure}[htbp]
    \centering
    \includegraphics[width=\columnwidth]{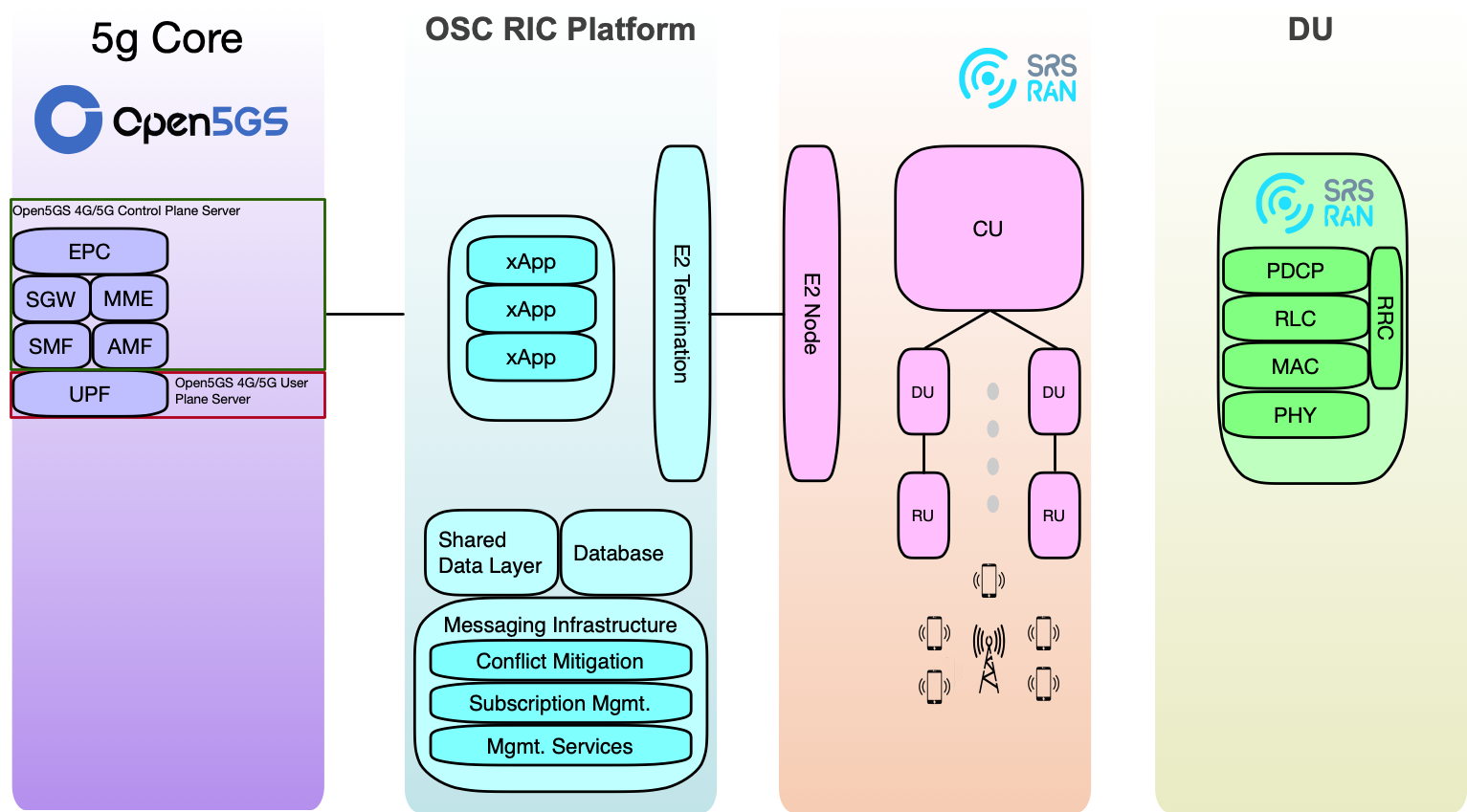}
    \caption{ORAN Architecture Overview: Integrating the 5G Core with the OSC RIC Platform and DU (Distributed Unit), showcasing modularity and support for xApp development.}
    \label{fig:oran_architecture}
\end{figure}

\subsection{AI/ML xApps in ORAN}
Artificial intelligence (AI) and machine learning (ML) are central to ORAN’s evolution, particularly through the development of xApps—intelligent applications that run on the RAN Intelligent Controller (RIC). Melodia et al. \cite{Kumar2021} introduce xDevSM, a flexible framework for streamlining xApp development within ORAN’s E2 interface. This framework facilitates rapid prototyping and deployment of AI/ML models, enhancing real-time functionalities such as traffic prioritization, resource scheduling, and interference management. By dynamically influencing network behavior, xApps enable ORAN to respond to changing conditions with greater agility and precision.

In addition to improving operational efficiency, AI/ML-based xApps significantly enhance resource allocation and energy efficiency within ORAN environments. Wang et al. \cite{Wang2021} demonstrate how embedded machine learning models within xApps optimize VNF splitting decisions, allowing for more effective management of distributed resources across ORAN nodes. This optimization not only reduces operational costs but also improves the overall performance of the network, making it more adaptable to varying demands.

The proactive nature of AI/ML in ORAN extends to network resilience, as explored by Allen et al. \cite{Allen2021}. Their research highlights how AI-driven xApps can predict, detect, and mitigate potential issues before they disrupt network services. By embedding these self-healing capabilities directly into the network, ORAN becomes more robust, minimizing downtime and ensuring consistent service delivery even in the face of misconfigurations or unforeseen challenges.

Unique use cases for AI/ML-driven xApps are exemplified in the AERPAW testbed. Mandal et al. \cite{Mandal2022} demonstrate the application of xApps in rapidly changing UAV scenarios, where ORAN’s adaptability is critical for managing unpredictable conditions. These experiments highlight the potential of xApps to extend ORAN’s applicability to novel domains, including emergency response operations and autonomous vehicle networks. The success of these implementations underscores ORAN’s ability to foster resilient, scalable, and innovative wireless communication solutions.

\section{Conclusions}
The advancement of wireless communication technologies is increasingly driven by the need for adaptable, efficient, and intelligent solutions. This reading report has analyzed pivotal research contributions in Dynamic Spectrum Sensing (DSS) and Open Radio Access Networks (ORAN), emphasizing the transformative role of Artificial Intelligence (AI) and Machine Learning (ML) in addressing the complexities of next-generation wireless systems. By leveraging AI-driven methodologies, these technologies have set new benchmarks for network scalability, adaptability, and performance.

In DSS, foundational frameworks like DeepSense have demonstrated the feasibility of integrating deep learning into real-time spectrum sensing, significantly improving accuracy and efficiency. Subsequent advancements, including DeepSweep and Wideband Signal Stitching, addressed limitations in latency, scalability, and dataset diversity. These efforts have not only enhanced the robustness of spectrum sensing solutions but also expanded their applicability to real-world, latency-sensitive applications such as autonomous vehicles, industrial automation, and urban IoT networks.

ORAN’s progression has similarly been marked by innovation, characterized by its open, modular architecture and the incorporation of AI/ML to enhance network intelligence. Digital twins have emerged as critical tools for optimizing network configurations and ensuring operational resilience, while experimental platforms like AERPAW have validated ORAN’s adaptability in dynamic scenarios. Furthermore, AI/ML-driven xApps have introduced real-time decision-making, predictive analytics, and self-healing capabilities, enabling ORAN to address diverse use cases such as UAV communication, smart cities, and emergency response.

Looking ahead, future research must focus on overcoming remaining challenges, such as reducing computational overhead, enhancing real-time adaptability, and scaling these frameworks to support higher-frequency bands, including terahertz. Promising directions include the integration of reinforcement learning for autonomous spectrum optimization, edge computing for distributed processing, and the development of scalable AI architectures tailored for next-generation networks. These advancements will be instrumental in creating intelligent, self-optimizing communication networks capable of meeting the demands of 5G, 6G, and beyond.

By synthesizing insights from foundational and emerging research, this report underscores the pivotal role of AI/ML in reshaping the wireless communication landscape. Through continuous innovation, these technologies are poised to drive transformative advancements across industries, delivering unprecedented levels of connectivity, efficiency, and reliability.

\nocite{*}
\bibliographystyle{IEEEtran}
\bibliography{References}

\end{document}